\documentclass[
 reprint,
 nofootinbib,
 amsmath,amssymb,
 aps,
]{revtex4-2}

\usepackage{graphicx}
\usepackage{dcolumn}
\usepackage{bm}
\usepackage{xcolor}
\usepackage{amsmath}
\usepackage{mathrsfs}
\usepackage{amssymb}
\usepackage{braket}
\usepackage{tabularx}

\begin{document}


\title{Optimizing the incident electron momentum for resonant few-photon\\ 
Kapitza-Dirac scattering in bichromatic laser fields}

\author{Ingo~Elsner and Carsten~M\"uller}
\address{Institut f\"ur Theoretische Physik I, Heinrich-Heine-Universit\"at D\"usseldorf, Universit\"atsstra{\ss}e 1, 40225 D\"usseldorf, Germany}
\date{\today}


\date{\today}

\begin{abstract}
Nonrelativistic Kapitza-Dirac scattering of electrons from counterpropagating bichromatic laser waves is studied in the resonant Bragg regime, taking the electron spin into account. We show that the intrinsic field-induced detuning, which arises in the Rabi oscillation dynamics between initial and scattering state of the electron, can be compensated by a suitable adjustment of its incident momentum. Analytical formulas of the optimized electron momentum for spin-dependent three-photon and spin-independent four-photon Kapitza-Dirac scattering are obtained from simplified model systems in reduced dimensionality, which preserve the characteristic properties of the process.
\end{abstract}

\maketitle


\section{Introduction}
The Kapitza-Dirac effect (KDE) refers to the quantum mechanical
scattering of an electron on the periodic potential formed 
by two counterpropagating laser waves \cite{KD,Review1,Review2}. 
It resembles the diffraction of light on a grating, with the roles 
of light and matter interchanged. In its original version \cite{KD}, 
the effect involves two photons from a standing wave: The electron 
absorbs a photon of momentum $\hbar k$ and emits another photon of 
momentum $-\hbar k$ (stimulated Compton scattering). In result, 
the electron is elastically scattered, reverting its longitudinal 
momentum from $-\hbar k$ to $\hbar k$. The momentum transfer from 
the standing wave to the electron thus amounts to $2\hbar k$.

This basic KDE version was experimentally observed in 2002 \cite{Freimund-PRL}
and is nowadays often referred to as two-photon KDE. It belongs to 
the Bragg regime of the process where the applied laser intensities 
are rather low and the transition between initial and final electron 
state proceeds {\it resonantly}. Also the complementary diffraction 
regime of KDE has been observed \cite{Freimund-Nature}, where a very 
broad momentum distribution in the final state results. Related 
experiments studied inelastic ponderomotive scattering of electrons
at high intensity \cite{Hommelhoff} and the KDE of atoms \cite{atoms}. 
Very recently, an ultrafast KDE has been observed in a free electron 
wave packet that was generated by atomic tunneling ionization in a 
standing laser wave \cite{ultrafast}.

Kapitza-Dirac scattering in the Bragg regime can also occur in bichromatic 
laser waves, where the process involves the exchange of more than two 
photons with the field \cite{Smirnova}. For example, in the three-photon KDE (3KDE),
an incident electron of longitudinal momentum $-2\hbar k$ interacts with 
a laser wave and its counterpropagating second harmonic, absorbing two 
'small' photons $\hbar k$ and emitting one 'big' photon $-2\hbar k$, 
resulting in a momentum transfer of $4\hbar k$ to the scattered electron. 

Interestingly, the electron spin degree of freedom can play a crucial
role in 3KDE, when the interaction arises from an ${\vec{A}}^2$ term in the 
Hamiltonian, in combination with a $\vec{\sigma}\cdot \vec{B}$ term. Then, an incident 
electron with, say, spin up can be scattered into a spin-down state 
\cite{Batelaan-2003,McGregor-2015,Dellweg-2016}. This spin-flipping transition 
can be controlled by the applied laser polarizations \cite{Dellweg-2017} and 
even allows to construct a spin-polarizing beam splitter for free electrons 
\cite{Dellweg-PRL}. The latter are also possible for the two-photon KDE when 
laser waves of circular polarization are used \cite{Bauke-2015, Ahrens-2017}.
Related studies have proposed spin polarization of free electrons by 
magnetic gratings \cite{McGregor-2011, Tang, lattice1}, space-variant
Wien filters \cite{Karimi}, optical lattices \cite{lattice2} and 
near-fields around nanowires \cite{near-field}.

We note that the spin-dependent 3KDE has originally been studied in a 
relativistic setting, where a high-energy electron is incident under a 
small angle on a standing laser wave \cite{Ahrens-2012}. In the electron 
rest frame, the comoving wave is red-shifted, whereas the counterpropagating 
wave is blue-shifted. Spin effects in KDE with relativistic electrons have 
been studied thoroughly in recent years, including beam shape and focusing 
effects \cite{Ahrens-2020, Ahrens-2022, Ahrens-2024}. In connection it is 
noteworthy that, at very high field intensities, ultrarelativistic electrons 
can also be spin-polarized radiatively by interactions with single \cite{Hatsagortsyan-2019, Hatsagortsyan-2020, Hatsagortsyan-2023} or counterpropagating \cite{Sorbo-2018, Seipt-2019} laser pulses. 

\begin{figure}[ht]
\includegraphics[width=0.4\textwidth]{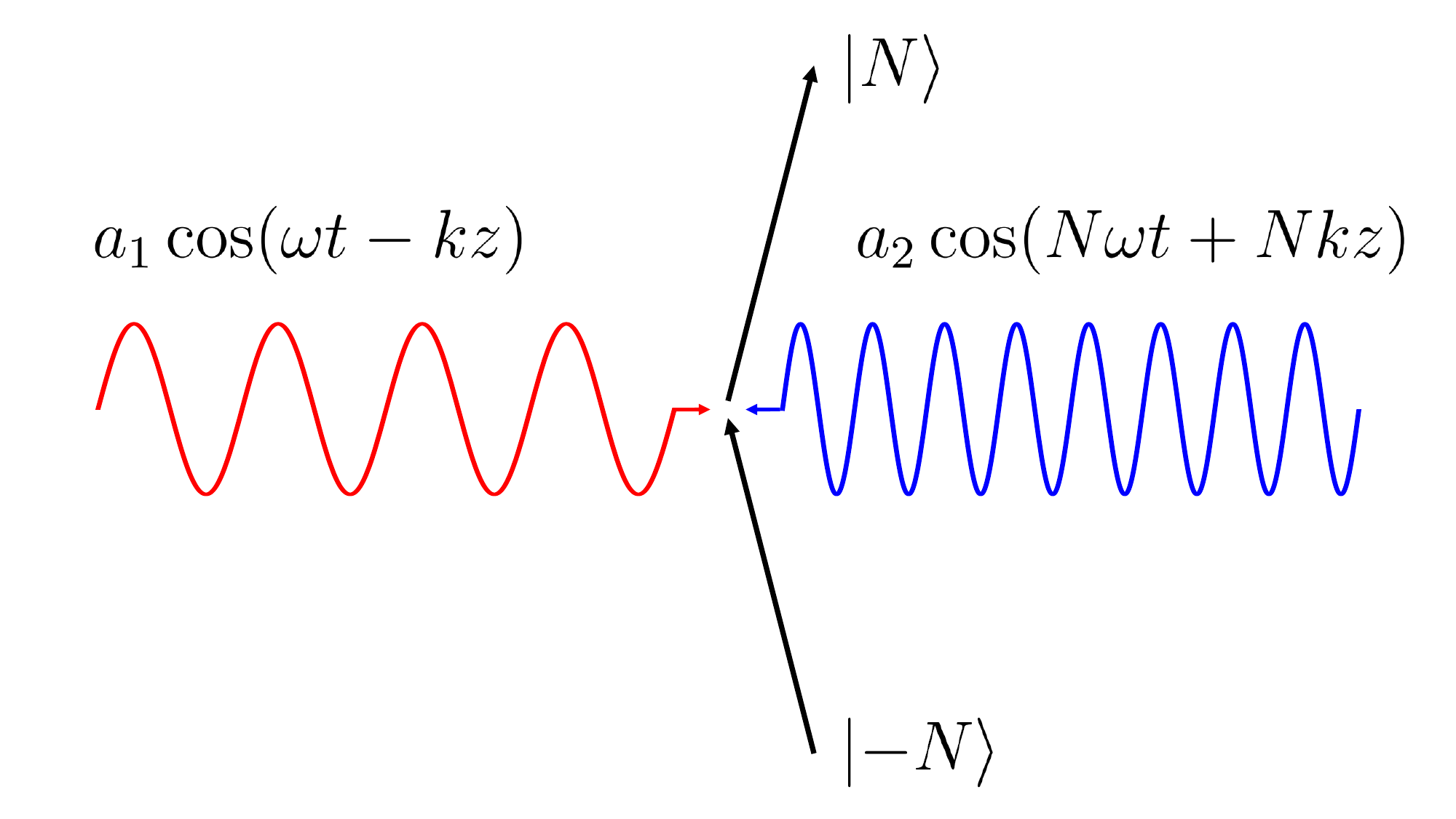}
\caption{Scheme of the $(N+1)$-Kapitza-Dirac effect in two counterpropagation laser waves with frequency $\omega$ (red) and $N\omega$ (blue) for the incident and outgoing electron states $\ket{-N}$ and $\ket{N}$, respectively (black arrows).}
\label{fig1_Intro}
\end{figure} 

\begin{figure}[ht]
\includegraphics[width=0.45\textwidth]{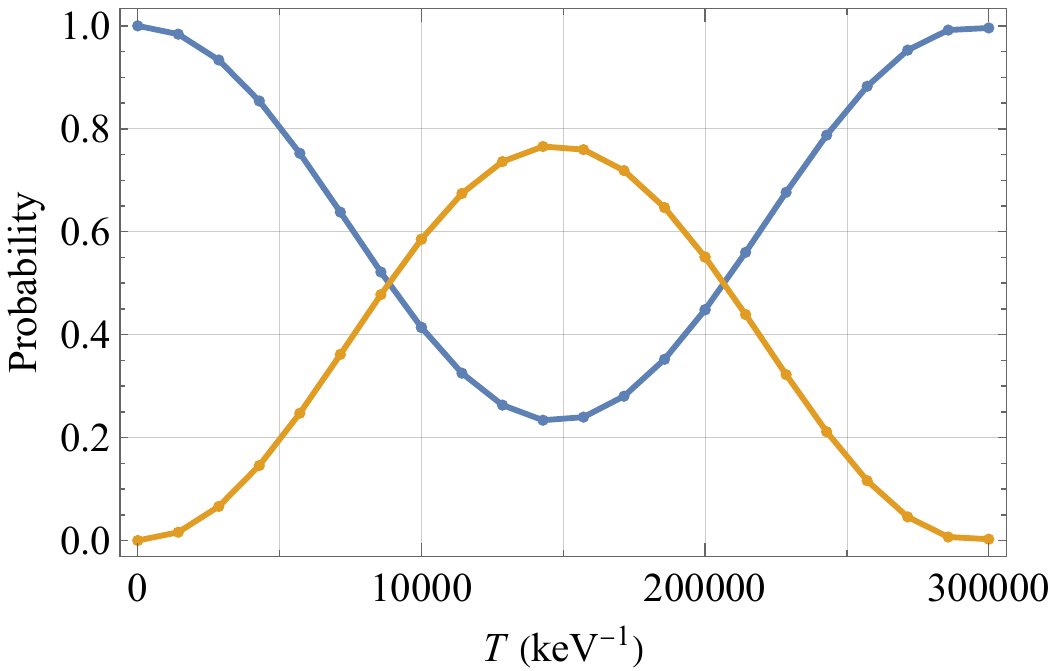}
\caption{Example of a detuned Rabi oscillation dynamics between initial electron state with incident momentum $-2 \hbar k$ and scattered state with momentum $2 \hbar k$. Shown are the final occupation probabilities $|c^\uparrow_{-2}(T)|^2$ (blue line) and $|c^\downarrow_{2}(T)|^2$ (orange line) for Kapitza-Dirac scattering from bichromatic counterpropageting laser waves as a function of the interaction time $T$. The field parameters are chosen as $|e| a_1 = 10$ keV, $|e| a_2 = 10$ keV and $\hbar \omega = 5$ keV [see Eq.~(\ref{vectorpotential})] and the incident electron offset is $p_z = 0$ [see Eq.~(\ref{ansatz})].}
\label{fig2_Detuning}
\end{figure}

Controlling and manipulating the spin of free electrons through Kapitza-Dirac
scattering offers very interesting opportunities. There exists an inherent 
problem, though: As a resonant process, the 3KDE proceeds most efficiently
when the energy-momentum balance of the electron is exactly satisfied. 
However, what matters are not the asymptotic electron momenta outside the 
interaction region, but instead the electron momenta {\it inside} the 
laser fields where the scattering event takes place. An electron with 
incident momentum $-2\hbar k$ therefore only seems to couple exactly resonantly
to the laser fields, but actually its momentum is slightly detuned. This 
intrinsic field-induced detuning leads to incomplete spin transitions 
in the 3KDE \cite{Dellweg-2016, Dellweg-2017}, as illustrated in Fig.~\ref{fig2_Detuning}. This detrimental phenomenon 
sets limitations on the efficiency of Kapitza-Dirac scattering as a means 
to purposefully influence the electron dynamics. We note that, while the 
field-dressed momentum of an electron inside a single laser wave is known 
analytically \cite{Volkov}, it is generally unknown in counterpropagating waves. For the latter case, only approximate solutions to the electron quantum dynamics exist \cite{Hu-2015,King-2016}, while the classical motion is highly nonlinear and can even become chaotic in strong fields \cite{Bauer-1995, Kaplan-2005, Lehmann-2012, Bashinov-2015, Lv-2021}.

The goal of the present paper is to analyze the field-induced detuning
in bichromatic Kapitza-Dirac scattering of nonrelativistic electrons. 
Our focus shall lie on the spin-dependent 3KDE, but for comparison we will 
also treat the spin-independent four-photon KDE (4KDE). Since a 
solution of the problem in full glory turns out to be very difficult, we 
shall consider simplified model systems which are obtained by suitable 
dimensional reductions. This will enable us to derive analytical formulas 
for those incident electron momenta which lead to exactly resonant scattering 
and thus cure the detuning problem. They will be called "optimum-$p$ formulas".

Before moving on we note that various interesting aspects of spin-independent 
Kapitza-Dirac processes have been studied, as well. They include relativistic 
treatments based on the Klein-Gordon equation \cite{Avetissian-1975, Fedorov},
electron scattering in dielectric media \cite{Avetissian-1976, Hayrapetyan}
and optical near fields \cite{Ropers}, multiple-pathway interferences 
\cite{Dellweg-2015, Vidil} and interferometric setups \cite{Marzlin}, as
well as high-intensity \cite{Fritzsche} and higher-order effects 
\cite{Kozak} in standing light fields.

Our paper is organized as follows. In Sec.~\ref{Analytic} we present the general framework for our KDE treatment and decompose the full problem into subsystems and associated channels of reduced dimensionality. Taking a particular channel as illustrative example, we show in Sec.~\ref{pFormula} how the detrimental detuning in the 3KDE spin dynamics can be avoided by either optimizing the incident electron momentum or the applied laser field parameters. The consideration is generalized to a generic 3KDE channel in Sec.~\ref{chapterGeneralSolution}, where also the corresponding transition probability, Rabi frequency and effective Hamiltonian are derived. In Sec.~\ref{GoingToEntireSystem} we argue how the channel results can be lifted to obtain an optimum-$p$ formula for the entire system and demonstrate the effectiveness of the formula by numerical simulations. The spin-preserving 4KDE is briefly discussed in Sec.~\ref{Treatment4KDE}, while conclusion and outlook are given in Sec.~\ref{ConclusionOutlook}.

For the sake of compact formulas, the reduced Planck constant $\hbar$ and the speed of light $c$ will be set to one.

\section{Theoretical Framework}\label{Analytic}

\subsection{System of differential equations}

We consider the time-dependent Pauli equation
\begin{align} \label{pauliEq}
i \frac{\partial}{\partial t} \psi = \frac{1}{2m} (\hat{\vec{p}} - e \vec{A})^2 \psi - \frac{e}{2m} \vec{\sigma} \cdot \vec{B} \psi
\end{align}
where $m$ is the electron mass, $e$ its charge and $\psi$ the electron spinor wave function. We denote the vector potential of the laser field with $\vec{A}$, which leads to the corresponding magnetic field $\vec{B} = \vec{\nabla} \times \vec{A}$. It couples to the electron spin involving the Pauli matrices $\vec{\sigma} = (\sigma_x, \sigma_y, \sigma_z)$. 

In order to model a general $\mathcal{N}$-KDE (i.e., a Kapitza-Dirac effect with exchange of $\mathcal{N}$ photons) we consider a laser wave with fundamental frequency $\omega$ and a counterpropagating laser wave of the $N$-th harmonic. The vector potential correspondingly reads
\begin{align}
\vec{A} = f(t) \hat{e}_x [a_1 \cos(\omega t - kz) + a_2 \cos(N \omega t + N k z)] \label{vectorpotential}
\end{align}
with $N = \mathcal{N} - 1$, the amplitudes $a_1$ and $a_2$, and the polarisation along the $x$ axis. The wave vector for the first field is $\vec{k} = k \hat{e}_z$ and for the second field $\vec{k}' = - N k \hat{e}_z$. We also introduce an envelope function $f(t)$, allowing to switch the field on and off after the interaction time $T$ in numerical calculations.

As in Ref.\,\cite{Dellweg-2016} we choose the incident electron momentum to be in the $y$-$z$ plane, so that $\vec{p} \cdot \vec{A} = 0$ holds. Due to the periodicity of the laser field along $z$, we use an ansatz in the form of a discrete plane-wave expansion 
\begin{align} \label{ansatz}
\psi(z,t) = \sum_n c_n(t)\, e^{i (n k + p_z) z} = \sum_n c_n(t) \ket{n}
\end{align}
with spinorial coefficients $c_n = \begin{pmatrix}
c_n^{\uparrow}\\
c_n^{\downarrow}
\end{pmatrix}$ and a momentum offset $p_z$, which is assumed to be small, $p_z \ll k$. By properly adjusting its value, the detuning effect can be compensated, as will be shown below. The plane waves with momentum $n k + p_z$ are denoted as $\ket{n}$. We can rewrite Eq.~(\ref{pauliEq}) as a system of infinitly many coupled differential equations:
\begin{align}
i \dot{c}_n(t) = E_n c_n(t) + \mathcal{V}_n(t) + \mathcal{W}_n(t)
\end{align}
with kinetic energies $E_n = \frac{(nk + p_z)^2}{2m}$ and 
\begin{align} \label{MatthiasTurm}
&\mathcal{V}_n = f(t)^2  \Bigg[\frac{e^2 a_1^2}{8m}  \Big(e^{2 i \omega t} c_{n+2}(t)  + 2 c_n(t) + e^{-2 i \omega t} c_{n-2}(t)  \Big) \notag\\
&+ \frac{e^2 a_1 a_2}{4m}  \Big(e^{-i(N+1) \omega t} c_{n+N-1}(t) +  e^{i(N+1) \omega t} c_{n-N+1}(t) \notag\\
&+e^{-i(N-1)\omega t} c_{n+N+1}(t) + e^{i(N-1)\omega t} c_{n-N-1}(t) \Big) \notag\\
&+ \frac{e^2 a_2^2}{8m}  \Big(e^{-2i N \omega t} c_{n+2N}(t)  + 2 c_n(t) + e^{2 i N \omega t} c_{n-2N}(t)    \Big)\Bigg],\notag\\
&\mathcal{W}_n = f(t)  \sigma_y \Bigg[\frac{i e \omega a_1}{4m} \Big(e^{i \omega t} c_{n+1}(t) - e^{- i \omega t} c_{n-1}(t)    \Big)  \notag \\
&+ \frac{i e \omega N a_2}{4m} \Big( e^{- i N \omega t} c_{n+N}(t) - e^{i N \omega t} c_{n-N}(t)    \Big) \Bigg].
\end{align}
In the following, we always assume that the initial state $\ket{-N}$ is fully populated with spin up at $t = 0$, so that our initial conditions are $c_{-N}^\uparrow(0) = 1$, $c_{-N}^\downarrow(0)=0$, and $c_n^{\uparrow,\downarrow}(0) = 0$ for all $n \neq -N$.

We note that, since the laser field is independent of $y$, the incident electron momentum component $p_y$ is conserved and has thus been dropped from the ansatz in Eq.~\eqref{ansatz}; the corresponding contribution of $p_y$ to the electron energy can be removed via a phase transformation.

\subsection{Transition potentials} \label{combinatoricNkde}

Apart from numerical calculations, an analytical treatment of the 3KDE was performed in Ref.~\cite{Dellweg-2016} within time-dependent perturbation theory. In such an approach, however, the detuning phenomenon does not become evident, rendering a nonperturbative approach necessary. Nevertheless from the corresponding consideration we can gain insights on how the KDE evolves by identifying the relevant transition potentials in Eq.~\eqref{MatthiasTurm} which can induce transitions between the initial state $\ket{-N}$ and final state $\ket{N}$ through a $\mathcal{N}$-KDE in leading order. These potentials are
\begin{align}
V_1(t) &= \frac{e^2 a_1^2}{8m} e^{-2 i \omega t} \sum_n \ket{n} \bra{n+2}\,,\nonumber\\
V_2(t) &= \frac{e^2 a_1 a_2}{4m} e^{-(N-1) i \omega t} \sum_n \ket{n} \bra{n+N+1}\,,\nonumber\\
W_1(t) &= \frac{i e a_1 \omega}{4m} \sigma_y  e^{i \omega t} \sum_n \ket{n} \bra{n+1}\,,\nonumber\\
W_2(t) &= \frac{i e  a_2 N \omega}{4m} \sigma_y e^{-N i \omega t} \sum_n \ket{n} \bra{n+N}\,.
\end{align}
Suitable combinations of them can lead to transitions in leading order, while the remaining two potentials 
\begin{align}
V_3(t) &= \frac{e^2 a_1 a_2}{4m} e^{-i (N+1) \omega t} \sum_n \ket{n} \bra{n+1}\,,\nonumber\\
V_4(t) &= \frac{e^2 a_2^2}{8m} e^{-2 i  \omega t} \sum_n \ket{n} \bra{n+2N}
\end{align}
are only capable of inducing transitions in higher orders when combined with other potentials.

Because of their particular spin properties, KDE processes with uneven $\mathcal{N}$ are of special interest to us. For those cases the important potentials $V_1$, $V_2$, $W_1$ and $W_2$ are to be combined in a specific way to obtain the required momentum transfer of $2Nk$ along with a Rabi frequency for the $\mathcal{N}$-KDE of the form $\Omega_{\mathcal{N}\text{-KDE}} \propto a_1^N a_2 \omega$. If $W_2$ is present, it is mandatory to combine solely $V_1$ in the process $N/2$ times. If instead $W_1$ is used, $V_2$ has to be utilized once and then $V_1$ has to be present $N/2-1$ times. In total we need $N/2 + 1$ potentials for a $\mathcal{N}$-KDE in leading order. We can see from this, that the effect takes place in two separate manners, which we will call ``subsystems''. For subsystem I involving $W_2$ we obtain $N$ different combinatorial possibilities how the effect can be realized, which we will call ``channels''. For subsystem II where $W_1$ is involved, there are $N (N-1)$ resulting channels for the effect. Typical examples for $\mathcal{N}$-KDE channels would accordingly be
\begin{align*}
&\text{subsystem I:}\ \ \ c_{-N}^\uparrow \overset{W_2}{\longrightarrow} c_0^\downarrow  \overset{V_1}{\longrightarrow} c_2^\downarrow \overset{V_1}{\longrightarrow}... \overset{V_1}{\longrightarrow} c_N^\downarrow \\
&\text{subsystem II:}\ \ c_{-N}^\uparrow \overset{W_1}{\longrightarrow} c_{-N+1}^\downarrow  \overset{V_2}{\longrightarrow} c_2^\downarrow \overset{V_1}{\longrightarrow}... \overset{V_1}{\longrightarrow} c_N^\downarrow
\end{align*}
We note that KDE processes with even $\mathcal{N}$ do not consist of two separate subsystems, because there is only the combination of $V_2$ with several instances of $V_1$. We will briefly analyze this scenario for the 4KDE in Sec.\,\ref{Treatment4KDE}.

\subsection{Minimal model systems for 3KDE}

After having established that the $\mathcal{N}$-KDE can be decomposed into two separate subsystems with their specific channels, we can analyze how this can help us to treat the 3KDE in a nonperturbative way. For the 3KDE we have subsystem I employing $V_1$ and $W_2$ with two channels, while subsystem II employs $V_2$ and $W_1$ having again two channels, where a channel refers to the order in which the interactions occur:
\begin{align}
&\text{subsystem I:} \notag\\
&\quad V_1(t) = \frac{e^2 a_1^2}{8m} e^{2 i \omega t} \sum_n \ket{n} \bra{n+2}\\
&\quad W_2(t) = \frac{i e a_2 \omega }{2m}  \sigma_y e^{-2 i \omega t} \sum_n \ket{n} \bra{n+2}\\
&\quad \text{channel 1:} \quad c_{-2}^\uparrow \overset{V_1}{\longrightarrow} c_{0}^\uparrow \overset{W_2}{\longrightarrow} c_{+2}^\downarrow \label{sub1channel1zustaende}\\
&\quad \text{channel 2:} \quad c_{-2}^\uparrow \overset{W_2}{\longrightarrow} c_{0}^\downarrow \overset{V_1}{\longrightarrow} c_{+2}^\downarrow \label{sub1channel2zustaende} 
\end{align}
\begin{align}
&\text{subsystem II:} \notag\\
&\quad V_2(t) = \frac{e^2 a_1 a_2}{4m} e^{- i \omega t} \sum_n \ket{n} \bra{n+3}\\
&\quad W_1(t) = \frac{i e a_1 \omega}{4m} \sigma_y e^{ i \omega t} \sum_n \ket{n} \bra{n+1}\\
&\quad \text{channel 1:} \quad c_{-2}^\uparrow \overset{V_2}{\longrightarrow} c_{+1}^\uparrow \overset{W_1}{\longrightarrow} c_{+2}^\downarrow\\
&\quad \text{channel 2:} \quad c_{-2}^\uparrow \overset{W_1}{\longrightarrow} c_{-1}^\downarrow \overset{V_2}{\longrightarrow} c_{+2}^\downarrow
\end{align}

If we restrict our consideration to subsystem I, for example, then Eq.~\eqref{MatthiasTurm} becomes considerably smaller 
\begin{align} \label{red-sys}
i \dot{c}_n(t) &= [E_n +2(e^2 a_1^2 + e^2 a_2^2)] c_n(t) \notag\\
&\quad + \frac{e^2 a_1^2}{8m} [e^{-2i \omega t} c_{n-2}(t)+ e^{2i \omega t} c_{n+2}(t)] \notag\\
& \quad + \frac{i e a_2 \omega}{2m} \sigma_y  [e^{2i\omega t} c_{n-2}(t) + e^{-2i\omega t}  c_{n+2}(t)]\,,
\end{align}
where we have set the envelope function to $f(t)=1$.
Even though the system is now more concise, it still contains infinitely many coupled differential equations and therefore not only describes processes of leading order but also transitions of higher order. To isolate the KDE in lowest order and to render it accessible to an analytical treatment, we have to further reduce Eq.~\eqref{red-sys} to a minimal model system, whose extent is bounded by the initial and final state. For subsystem I, this leads us to the following truncated matrix equation:
\begin{align}
i \dot{\vec{c}}_\text{I}(t) &= \begin{pmatrix}
E_{-2}& \tilde{V}_1 &\tilde{W}_2 &0\\
\tilde{V}_1^*& E_0 &0 &\tilde{W}_2\\
\tilde{W}_2^*& 0 &E_0 &\tilde{V}_1\\
0& \tilde{W}_2^* &\tilde{V}_1^* &E_2
\end{pmatrix} \vec{c}_\text{I}(t) \label{matrixSub1}
\end{align}
with $\vec{c}_\text{I}(t) = (c^{\uparrow}_{-2}(t),c^{\uparrow}_0(t),c^{\downarrow}_0(t),c^{\downarrow}_2(t))$ and the potentials
\begin{align}
\tilde{V}_1 &= \frac{e^2 a_1^2}{8m}e^{2 i \omega t}, \quad \tilde{W}_2 = \frac{e a_2 \omega}{2m}e^{-2 i \omega t}\,.
\end{align}
A constant term $2(e^2 a_1^2 + e^2 a_2^2)$ has been removed here from the diagonal entries via a phase transformation.

Equivalently, we can construct a minimal model for subsystem II as
\begin{align}
i \dot{\vec{c}}_\text{II}(t) &= \begin{pmatrix}
E_{-2}& \tilde{V}_2 &\tilde{W}_1 &0\\
\tilde{V}_2^*& E_{1} &0 &\tilde{W}_1\\
\tilde{W}_1^*& 0 &E_{-1} &\tilde{V}_1\\
0& \tilde{W}_1^* &\tilde{V}_2^* &E_2
\end{pmatrix} \vec{c}_\text{II}(t)
\end{align}
with $\vec{c}_\text{II} = (c^{\uparrow}_{-2}(t),c^{\uparrow}_{1}(t),c^{\downarrow}_{-1}(t),c^{\downarrow}_2(t))$ and
\begin{align}
\tilde{V}_2 &= \frac{e^2 a_1 a_2}{4m}e^{- i \omega t}, \quad \tilde{W}_1 = \frac{e a_1 \omega}{4m}e^{ i \omega t}\,.
\end{align}
We are able to treat these minimal model systems analytically and derive formulas for the momentum offset $p_z$ so that the detuning effect within the KDE is cured.

\section{Analyzing channel 1 of subsystem~I}\label{pFormula}

The easiest way to approach the subsystems analytically is by treating a single channel. We will consider for subsystem I the channel 1 as an example [see Eq.~(\ref{sub1channel1zustaende})], which has the minimal model system
\begin{align} \label{matrixsub1channel1}
i \dot{\vec{c}}_\text{I}(t) = \begin{pmatrix}
E_{-2}& \tilde{V}_1 &0 &0\\
\tilde{V}_1^*& E_0 &0 &\tilde{W}_2\\
0& 0 &E_0 &0\\
0& \tilde{W}_2^* &0 &E_2
\end{pmatrix} \vec{c}_\text{I}(t)\,.
\end{align}
In comparison with Eq.~(\ref{matrixSub1}), the coupling of $c_{-2}^\uparrow$ and $c_0^\downarrow$ via $\tilde{W}_2$ has been removed as well as the coupling of $c_0^\downarrow$ and $c_2^\downarrow$ via $\tilde{V}_1$, because they are associated with the channel 2 [see Eq.~(\ref{sub1channel2zustaende})]. We see that in this scenario the system is forced to transition to the intermediate state $c_0^\uparrow$ and that its counterpart $c_0^\downarrow$ is redundant. Therefore the system reduces even more to only three coupled equations which we can analyse and solve. To do so we take the first and last equation of Eq.~\eqref{matrixsub1channel1} and solve for the intermediate state and insert separately into the middle equation. This protocol leads us to a set of two coupled equations, which do not contain any exponential terms anymore:
\begin{align} \label{sub1channel1secondOrderEq}
&\ddot{c}_{-2}^\uparrow(t) + i (E_{-2} + E_0 - 2 \omega)  \dot{c}_{-2}^\uparrow(t) \notag \\
&+ \left(\frac{e^4 a_1^4}{64m^2} - E_{-2}(E_0 - 2 \omega)\right) c_{-2}^\uparrow(t) + \frac{e^3 a_1^2 a_2 \omega}{16m^2} c_{2}^\downarrow(t) = 0 \nonumber\\
&\ddot{c}_{2}^\downarrow(t) + i (E_{2} + E_0 - 2 \omega)  \dot{c}_{2}^\downarrow(t) \notag \\
&+ \left(\frac{e^2 a_2^2 \omega^2}{4m^2} - E_{2}(E_0 - 2 \omega)\right) c_{2}^\downarrow(t)  + \frac{e^3 a_1^2 a_2 \omega}{16m^2} c_{-2}^\uparrow(t) = 0
\end{align}

The resulting equations represent a system of two harmonic oscillators which are coupled via a hook term $\frac{e^3 a_1^2 a_2 \omega}{16m^2}  c_{\pm 2}^{\downarrow \uparrow}$. Note that the quantity $\omega_c^2 = -\frac{e^3 a_1^2 a_2 \omega}{16m^2} > 0$ closely resembles the Rabi frequency $\Omega_\text{3KDE} = 8\omega_c^2 / m$ of the 3KDE \cite{Dellweg-2016}. 

\subsection{Optimum-p formula}

To achieve a complete transfer between the initial-state and scattered-state populations in a periodic manner requires $c_{-2}^\uparrow(t)$ and $c_2^\downarrow(t)$ to behave in a mutually symmetric way. This will happen if the system of equations \eqref{sub1channel1secondOrderEq} has a symmetric structure itself. Since $p_z \ll k$ the energies of the initial and final states are approximately equal,
\begin{align}
E_{-2} \approx E_2\,,
\end{align}
so that the system becomes symmetric, if the oscillator frequency terms are set equal. From the condition
\begin{align*}
\frac{e^4 a_1^4}{64m^2} - E_{-2}(E_0 - 2 \omega) \overset{!}{=} \frac{e^2 a_2^2 \omega^2}{4m^2} - E_{2}(E_0 - 2 \omega)
\end{align*}
it follows that
\begin{align}
- \frac{4 \omega p_z}{m} = E_{-2} - E_{2} = \frac{e^4 a_1^4 - 16 e^2 a_2^2 \omega^2}{64m^2 (E_0 - 2\omega)}
\end{align}
which implies the following optimum-$p$ formula for the momentum offset
\begin{align}
p_{\text{I},1} \approx \frac{e^4 a_1^4 - 16 e^2 a_2^2 \omega^2}{512 m \omega^2}\,. \label{pformelSub1Channel1}
\end{align}
We interpret $p_\text{I,1}$ as the momentum shift which is necessary to compensate the field-dressing caused by the interaction potentials $\tilde{V}_1$ and $\tilde{W}_2$ in the matrix of the channel Eq.~(\ref{matrixsub1channel1}). We can also read off a condition for optimal laser parameters, for which no detuning occurs for $p_z = 0$, which is
\begin{align}
e^4 a_1^4 = 16 e^2 a_2^2 \omega^2.
\end{align}
From our derivation we can see that this condition stems from the form of $\tilde{V}_1$ and $\tilde{W}_2$, and means that both potentials have to be equally strong.
 
\begin{figure}[ht]
\includegraphics[width=0.45\textwidth]{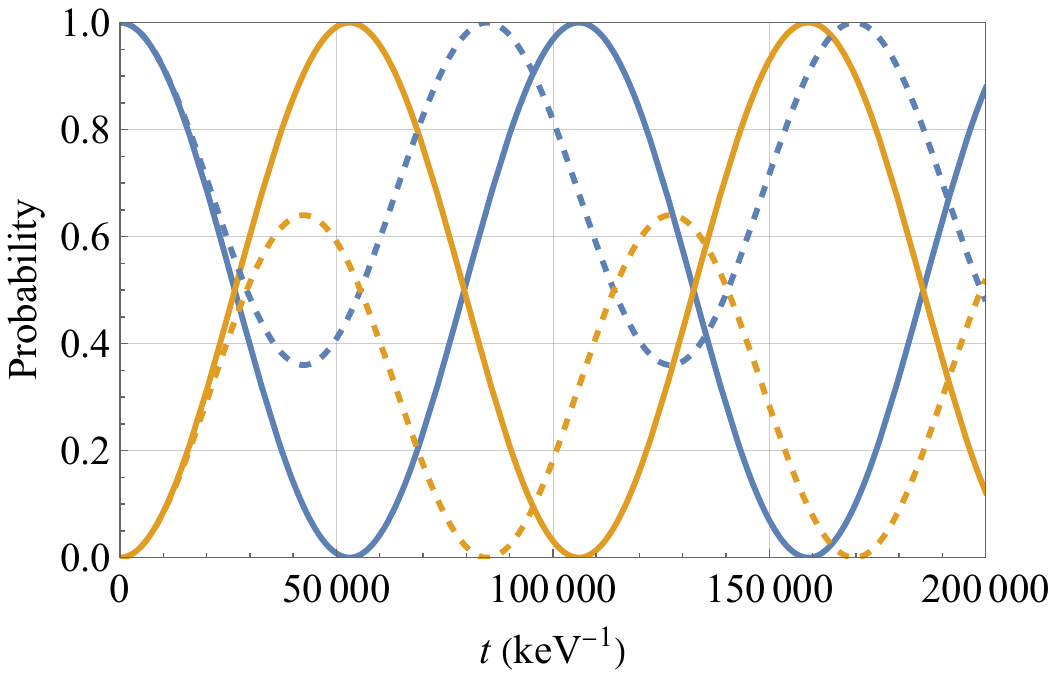}
\caption{Rabi oscillation of the occupation probabilities $|c_{-2}^\uparrow(t)|^2$ (blue lines) and $|c_2^\downarrow(t)|^2$ (orange lines)  of subsystem I channel 1 plotted against time $t$ for the parameters $|e| a_1 = 10$\,keV, $|e| a_2 = 2.5$\,keV, and $\omega = 5$\,keV. The dashed lines show strong detuning for $p_z = 0$, whereas the solid lines show a fully developed oscillation for $p_z \approx 0.0012$\,keV satisfying the optimum-$p$ formula \eqref{pformelSub1Channel1}.}
\label{fig3_Sub1Channel1}
\end{figure}

Figure~\ref{fig3_Sub1Channel1} shows the Rabi oscillation dynamics between the populations of the incident and scattered states for channel 1 of subsystem I in a case where the laser parameters are not optimally chosen ($|e| a_1 = 10$\,keV, $|e| a_2 = 2.5$\,keV, $\omega = 5$\,keV). A vanishing offset $p_z = 0$, represented by the dashed lines, leads to a strong detuning with the scattering probability $|c_{2}^\downarrow(t)|^2$ reaching a maximum of only $64 \%$. The solid lines refer to the same laser parameters and an offset $p_z \approx 0.0012$\,keV set to the value given by Eq.~(\ref{pformelSub1Channel1}). For this case the intrinsic field-induced detuning is compensated, leading to a complete population transfer between the states. 

We note at this point that, for reasons of computational feasibility, our numerical calculations assume intense laser fields with frequencies in the x-ray domain. Corresponding radiation sources are, in principle, available via high-harmonic generation from plasma surfaces \cite{plasmaHHG} or high-power x-ray free-electron lasers (XFELs) like the European XFEL (Hamburg, Germany) or the Linac Coherent Light Source (LCLS, Stanford, California) \cite{XFEL}, which are able to produce brilliant x-ray pulses with photon energies up to $\sim 10$\,keV at $\sim 10^{20}$\,W/cm$^2$ intensities and $\sim 100$\,fs pulse durations.

\subsection{Decoupling for optimal laser parameters}

For optimal laser parameters, where $p_{\text{I},1}=0$ and the oscillator frequency terms in the coupled system \eqref{sub1channel1secondOrderEq} are equal, we see that the system decouples. With the transformation $q_\text{1/2}(t) = c_{-2}(t)^\uparrow \pm c_{2}^\downarrow(t)$ the system becomes
\begin{align}
&\ddot{q}_1(t) + i B \dot{q}_1(t) + (\omega_s^2 - \omega_c^2) q_1(t) = 0\,,\nonumber \\
&\ddot{q}_2(t) + i B \dot{q}_2(t) + (\omega_s^2 + \omega_c^2) q_2(t) = 0
\end{align} 
with $B = E_{-2} + E_0 - 2 \omega =  E_0 + E_{2} - 2 \omega $, $\omega_s^2 = \frac{e^4 a_1^4}{64m^2} - E_{-2}(E_0 - 2 \omega)= \frac{e^2 a_2^2 \omega^2}{4m^2} - E_{2}(E_0 - 2 \omega)  $. Those decoupled harmonic oscillators are solved by
\begin{align}
q_1(t) &=   A_1 e^{i \omega_1^+ t} + B_1 e^{i \omega_1^- t}\\
q_2(t) &=   A_2 e^{i \omega_2^- t} + B_2 e^{i \omega_2^+ t}
\end{align}
with the eigenfrequencies
\begin{align}
\omega^\pm_{1/2} &= - \frac{B}{2} \pm  \sqrt{\frac{B^2}{4} + (\omega_s^2 \mp \omega_c^2)}
\end{align}
and the coefficients 
\begin{align}
&A_1 =  \frac{E_{-2} + \omega_1^-}{\omega_1^+ - \omega_1^+} \approx 1- B_1, \: B_1 =  \frac{E_{-2} + \omega_1^+}{\omega_1^+ - \omega_1^-} \approx 1\,, \nonumber \\
&A_2 =  \frac{E_{-2} + \omega_2^+}{\omega_2^+ - \omega_2^-} = 1, \: B_2 =  \frac{E_{-2} + \omega_2^-}{\omega_2^- - \omega_2^+} = 0.
\end{align}

Here we make an important observation: For the optimal case, the coefficient $B_2$ vanishes. This is due to the fact, that we initially had only three coupled equations of first order from which we moved on to two coupled equations of second order. By this we introduced a spurious solution, which has no physical meaning and is not realized by the system itself. The coefficients $B_1$ and $A_2$ correspond to the slow frequencies of the inital and final state of the system, whereas the small coefficient $A_1$ is associated with the fast oscillation of the intermediate state. The resulting Rabi oscillation occurs with a beat frequency between the two slow oscillations of the initial and final states. We can then express the solution of our original system as
\begin{align}
|c_{-2}^\uparrow(t)|^2 &= \Big|\frac{1}{2} \Big[ q_1(t) + q_2(t) \Big]\Big|^2 \nonumber\\
& \approx \frac{B_1^2}{4}  + \frac{A_2^2}{4} + \frac{B_1 A_2}{4}  \cos^2 \left( \frac{1}{2} (\omega_1^- - \omega_2^-) t  \right) \nonumber\\
&  = \cos^2 \left(\frac{1}{2} \Omega_{\text{I,1}}^{(0)} t \right)
\end{align}
and similarly for $|c_{2}^\downarrow(t)|^2$. The Rabi frequency contains no detuning in the present optimal scenario and results from
\begin{align}
\Omega_\text{I,1}^{(0)} = \omega_1^- - \omega_2^- \approx \frac{\omega_c^2}{\sqrt{ \frac{B^2}{4} + \omega_s^2 }}
\end{align}
which reads, if we consider only leading orders in $m$,
\begin{align}
\Omega_\text{I,1}^{(0)} 
\approx -\frac{m}{\omega(m + \omega)} \frac{e^3 a_1^2 a_2 \omega}{16m^2}\,. \label{sub1channel1freq}
\end{align}
We note that this Rabi frequency for channel 1 of subsystem I is by a factor $\sim\frac{m}{\omega}$ larger than the Rabi frequency of the entire subsystem or of the whole 3KDE \cite{Dellweg-2016}.

\section{Solution for a general channel} \label{chapterGeneralSolution}

\subsection{Optimum-p formula}

After having treated channel 1 of subsystem I we can generalize our analysis to an arbitrary channel of the 3KDE. Let us consider a general system
\begin{align}
i \dot{\vec{c}}(t) = \begin{pmatrix}
E_{-2} & \tilde{U}_1 &  0 \\
\tilde{U}_1^* & E_i &  \tilde{U}_2  \\
0 & \tilde{U}_2^* &  E_2
\end{pmatrix} \vec{c}(t)
\end{align}
with $\vec{c}(t)= (c_{-2}(t), c_i(t), c_2(t))$ and transition potentials
\begin{align}
\tilde{U}_1 = U_1 e^{i \omega_1 t}\,,\ \ \tilde{U}_2 = U_2 e^{i \omega_2 t}
\end{align}
where $\omega_1 + \omega_2 = 0$ holds and we have $\omega_1 = \kappa \omega$ with $\kappa \in \{\pm 1, \pm 2\}$. The energy of the intermediate state is $E_i$. For the case of channel 1 of subsystem I the potentials were $\tilde{U}_1 = \tilde{V}_1$ and $\tilde{U}_1 =\tilde{W}_2$ with $\omega_1 = 2 \omega$ and $E_i = E_0$. Then for any abstract channel the resulting second-order equations will look like
\begin{align}
&\ddot{c}_{-2}^\uparrow(t) + B_1 \dot{c}_{-2}^\uparrow(t) + A_1 c_{-2}^\uparrow(t) - \omega_c^2 c_2^\downarrow(t) = 0\,, \nonumber\\
&\ddot{c}_{2}^\downarrow(t) + B_2 \dot{c}_2^\downarrow(t) + A_2 c_2^\downarrow(t) - \omega_c^2 c_{-2}^\uparrow(t) = 0 \label{generalSystem2ndOrder}
\end{align}
with
\begin{align}
&A_1 = U_1^2 - E_{-2}(E_i - \omega_1)\,,\ B_1 = E_{-2} + (E_i - \omega_1)\,,\nonumber\\
&A_2 = U_2^2 - E_2(E_i + \omega_2)\,,\ B_2 = E_2 + (E_i + \omega_2)
\end{align}
and $\omega_c^2 = -U_1 U_2$.
Assuming again $E_{-2} \approx E_2$, we need the oscillator frequency terms to be equal for the system to be symmetric. From $A_1 \overset{!}{=} A_2$ we obtain
\begin{align} \label{energy-diff}
- \frac{4\omega p_z}{m} = E_{-2} - E_2 = \frac{U_1^2 - U_2^2}{E_i - \omega_1}\,.
\end{align}
Hence, to reach an optimized coupling, the difference between the energies of the initial and final state {\it outside} the field needs to attain a non-zero value, which can be considered as a result of the {\it energy dressing inside} the field. In terms of the required momentum offset, we obtain 
\begin{align} \label{p_g}
p_{\text{g}} \approx \frac{m}{4\kappa \omega}\frac{U_1^2 - U_2^2}{\omega}\,,
\end{align}
where $E_i \ll \omega_1$ was used. This relation implies that there is no detuning for $p_\text{g} = 0$, if $U_1 = U_2$.

\subsection{Interpreting the field-induced detuning from transition potentials}\label{interpretation}

To interpret our results it is insightful to just consider the initial and an intermediate state coupled by a single transition potential. The matrix for such a system reads
\begin{align}
i \dot{\vec{c}}(t) = \begin{pmatrix}
E_{-2} & U_1 e^{i \omega_1 t} \\
U_1 e^{-i \omega_1 t}  & E_{i}
\end{pmatrix} \vec{c}(t)
\end{align}
leading to the second-order equation
\begin{align}
&\ddot{c}_{-2}(t) + i (E_{-2}  + E_{i}  - \omega_1) \dot{c}_{-2}(t) \notag \\
& \quad + (U_1^2 - E_{-2} (E_i - \omega_1) ) c_{-2}(t) = 0\,.
\end{align}
We see that the coupling through the transition potential results in an oscillatory behavior of the initial state -- unless the oscillator frequency term vanishes, which is the case for
\begin{align}
E_{-2} = \frac{U_1^2}{E_i - \omega_1}.
\end{align}
This expression represents 'half of' Eq.~\eqref{energy-diff} and shows that the coupling potential $U_1$ leads to a dressing of the initial state energy $E_{-2}$. A similar analysis would yield the energy shift of the final state energy, corresponding to the second half of Eq.~\eqref{energy-diff}. We note that the energy dressing by $U_1^2/(E_i-\omega_1)$ shares certain similarities with the well-known ponderomotive energy shift in a plane laser wave $a\cos(\omega t-\vec{k}\cdot\vec{r}\,)$ which is $e^2a^2/(E-p_\parallel)$ \cite{Volkov}. 

While in Refs.~\cite{Dellweg-2016, Dellweg-PRL, Dellweg-2017} the detuning phenomenon of the 3KDE was attributed qualitatively to be field induced, our consideration offers a means towards its quantitative description and, thus, its elimination by a suitable choice of the momentum offset $p_{\text{g}}$. We see that the detuning arises from the energy contributions to the initial and final electron states caused by their interaction with the relevant parts of the laser potentials. Based on this insight we can expect that each interaction that couples to an electron state, will induce an associated energy shift.

\subsection{Channel eigenfrequencies}

In addition to our analysis of the optimum-$p$ formula, it is possible to formally solve the system of equations \eqref{generalSystem2ndOrder} and calculate the corresponding eigenfrequencies. To this end, we insert the ansatz $c_{-2}^\uparrow(t) = C_1 e^{i k t}$, $c_{2}^\downarrow(t) = C_2 e^{i k t}$ and can then write the system as
\begin{align}
\begin{pmatrix}
A_1 - B_1 k - k^2& - \omega_c^2 \\
  - \omega_c^2& A_2 - B_2 k - k^2
\end{pmatrix} \begin{pmatrix}
C_1\\
C_2
\end{pmatrix} = \begin{pmatrix}
0\\
0
\end{pmatrix}.
\end{align}
Setting the determinant to zero results in a characteristic equation of forth order, which is possible to solve, but in general cumbersome to handle. Fortunately, it decomposes here into a biquadratic equation, if $E_{-2} \approx E_2$ holds because then $B_2 \approx B_1$. It follows
\begin{align}
[A_1 - k (B_1 + k)][A_2 - k (B_1 + k)] -  \omega_c^4 = 0
\end{align}
with the four solutions
\begin{align} \label{eigenfreq}
k_{1/2}^{\pm} &= - \frac{B_1}{2} \notag\\
& \pm  \sqrt{\frac{1}{2} (A_1 + A_2) + \frac{1}{4} B_1^2 \mp \sqrt{\frac{1}{4} (A_1 - A_2)^2  +  \omega_c^4} }
\end{align}
where the superscript (subscript) refers to the $\pm$ sign outside (inside) the main square root.
From those four solutions, one describes the fast oscillation of the intermediate state, while the other fast one is spurious. For channel 1 (channel 2) of subsystem I, the fast oscillation is $k^+_2$ ($k^-_2$) while $k^+_1$ ($k^-_1$) is spurious, whereas for channel 1 (channel 2) of subsystem II, the fast oscillation is $k^-_2$ ($k^+_2$) while $k^-_1$ ($k^+_1$) is spurious. The other two solutions describe the slow oscillations of the initial and the scattered state; the Rabi oscillation between them results as a beat frequency from the difference between the slow eigenmodes.

\begin{figure}[ht]
\includegraphics[width=0.45\textwidth]{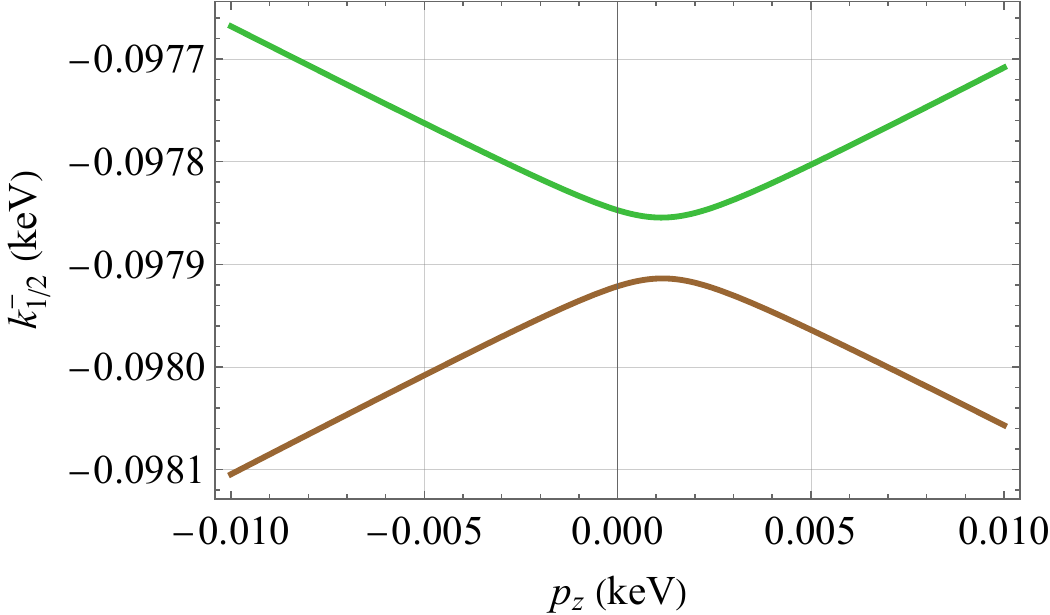}
\caption{Depiction of the eigenfrequencies $k_1^-$ (green line) and $k_2^-$ (brown line) of channel 1 in subsystem I as function of $p_z$ for the field parameters $|e|a_1 = 10$\,keV, $|e|a_2 = 2.5$\,keV, and  $\omega = 5$\,keV. An avoided crossing is seen whose point of closest approach is shifted from $p_z=0$ to the right due to unequal strengths of the interaction potentials [see Eq.~\eqref{p_g}].}
\label{fig4_AvoidedCrossing}
\end{figure}

The structure of Eq.~\eqref{eigenfreq} closely resembles the well-known energy eigenvalues of a coupled two-level system. The two slow eigenmodes, accordingly, exhibit a characteristic avoided crossing (or level repulsion) behavior \cite{Cohen-Tannoudji}. This is illustrated in Fig.~\ref{fig4_AvoidedCrossing} for the low eigenfrequencies $k_1^-$ and $k_2^-$ of channel 1 of subsystem I. For comparison we note that the fast mode (not shown) amounts to $k_2^+\approx 10$\,keV for the considered field parameters.

\subsection{Transition probabilities}

By including the initial conditions, we can obtain the full solution to the system \eqref{generalSystem2ndOrder}. From our analysis we already know that one of the four solutions is spurious; for channel 1 of subsystem I it is $k_1^+$. Hence, we can formulate the ansatz
\begin{align}
c_{-2}^\uparrow(t) = C_1^- e^{i k_1^- t} + C_2^- e^{i k_2^- t} + C_2^+ e^{i k_2^+ t} \label{ABsol}
\end{align}
and obtain, demanding $c_{-2}^\uparrow(0)=1$, the coefficients
\begin{align}
&C_1^- = \frac{U_1^2 + (E_{-2} + k_2^-)(E_{-2} + k_2^+)}{(k_1^- - k_2^-)(k_1^- - k_2^+)}\,,\nonumber\\
&C_2^- = \frac{U_1^2 + (E_{-2} + k_1^-)(E_{-2} + k_2^+)}{(k_2^- - k_1^-)(k_2^- - k_2^+)}\,,\nonumber\\
&C_2^+ = \frac{U_1^2 + (E_{-2} + k_1^-)(E_{-2} + k_2^-)}{(k_2^+ - k_1^-)(k_2^+ - k_2^-)}\,.
\end{align}
Because of the small difference $k_2^- - k_1^-$ in the denominator, the coefficients $C_1^-$ and $C_2^-$  are large compared to $C_2^+$, which is linked to the fast oscillating intermediate state. Therefore, we have $C_1^- + C_2^- \approx 1$ and can express the time evolution of the initial-state population as 
\begin{align}
|c_{-2}^\uparrow(t)|^2  \approx 1 -  4 C_1^- C_2^- \sin^2\left(\tfrac{1}{2} (k_2^- - k_1^-) t \right)
\end{align}

We show next that the quantity $4 C_1^- C_2^-$ has the shape of a Lorentz curve. To derive this, we notice that, on the one hand, the quantity 
\begin{align}
\Omega_\text{I,1}^{(\delta)} = k_1^- - k_2^-,
\end{align}
which appears in the denominator, represents the full Rabi frequency including the detuning (see Sec.~\ref{DeriveRabiFreq}). On the other hand, the remaining terms become the Rabi frequency without detuning [see Eq.~\eqref{sub1channel1freq}] after some expansions. First, we Taylor expand the frequencies $k_{1/2}^{-}$ by using $\sqrt{a+x} \approx \sqrt{a} + \frac{x}{2\sqrt{a}}$ for $x\ll a$. The remaining part in the denominator then becomes
\begin{align}
\frac{4}{(k_1^- - k_2^+)(k_2^- - k_2^+)} \approx \frac{m^2}{\omega^2 (m+\omega)^2}
\end{align}
which already represents the prefactor of the undetuned (squared) Rabi frequency. After some algebraic calculations the nominator can be written as
\begin{align}
&[U_1^2 + (E_{-2} + k_2^-)(E_{-2} + k_2^+)] \nonumber\\
&\quad \times [U_1^2 + (E_{-2} + k_1^-)(E_{-2} + k_2^+)] \approx - \omega_c^4
\end{align}
so that we arrive at the Lorentz curve
\begin{align}
4 C_1^- C_2^- &\approx \left(\frac{\Omega_\text{I,1}^{(0)}}{\Omega_\text{I,1}^{(\delta)}}\right)^{\!2} 
 =\frac{\left(\frac{e^3 a_1^2 a_2 \omega}{16m^2}\right)^2}{\left(\frac{e^3 a_1^2 a_2 \omega}{16m^2}\right)^2 + \delta^2_\text{I,1}}\,.
\end{align}
with the detuning parameter $\delta_\text{I,1}$ whose expression will be given in the next section. We note that for the final state population we then have
\begin{align}
|c_2^\downarrow(t)|^2 \approx 4 C_1^- C_2^- \sin^2\!\Big(\tfrac{1}{2}\Omega_\text{I,1}^{(\delta)}t\Big)
\end{align}

\subsection{Rabi oscillation with detuning} \label{DeriveRabiFreq}

As we have seen above, the Rabi oscillation results as a beat phenomenon between the slow eigenfrequencies of the initial and final state from Eq.~\eqref{eigenfreq}. By using the Taylor expansion $\sqrt{a + x} - \sqrt{a - x} \approx \frac{x}{\sqrt{a}}$, the Rabi frequency for a general channel can be expressed as
\begin{align}
\Omega_\text{g} &= \sqrt{\tfrac{\big[U_2^2 - U_1^2 + (E_2 - E_
{-2})(E_i - \omega_1)\big]^2 + 4 U_1^2 U_2^2}{2 \big( -U_1^2 - U_1^2 - (E_{-2} + E_{2}) (E_i - \omega_1) \big) +  (E_{-2} + E_i - \omega_1)^2}} \label{allgemeineFrequenz}
\end{align}
This formula can be rewritten more compactly as
\begin{align}
\Omega_\text{g} = \frac{1}{\varepsilon} \sqrt{(U_1 U_2)^2 + \delta_\text{g}^2}\,, \label{allgemeineFrequenzKurz}
\end{align}
with the general detuning parameter
\begin{align}
\delta_\text{g} = \frac{1}{2}\big[U_2^2 - U_1^2 + (E_2 - E_{-2})(E_i - \omega_1)\big]\,,
\end{align}
and the energy parameter
\begin{align}
\varepsilon &= \frac{1}{2}\Big[2 \big( -U_1^2 - U_1^2 - (E_{-2} + E_{2}) (E_i - \omega_1) \big)  \notag \\
& \quad +  (E_{-2} + E_i - \omega_1)^2\Big]^{1/2}\,. \label{energyparameter}
\end{align}
Accordingly, the Rabi oscillation with detuning reads
\begin{align}
|c_2^\downarrow(t)|^2 = \frac{1}{ 1+ \left( \frac{\delta_\text{g}}{ U_1 U_2} \right)^2}  
\sin^2\!\big(\tfrac{1}{2} \Omega_\text{g} t \big)\,. \label{transitionprobability}
\end{align}

For example, if we insert the corresponding potentials for channel 1 of subsystem I, we can expand the denominator in the frequency formula $\varepsilon$ [see Eq.~\eqref{allgemeineFrequenzKurz}] and reach a form of the Rabi frequency similar to Eq.~\eqref{sub1channel1freq}, but with the detuning incorporated:
\begin{align}
\Omega_{\text{I},1}^{(\delta)} 
&\approx \frac{m}{\omega ( \omega + m)} 
\sqrt{ \left(\frac{e^3 a_1^2 a_2 \omega}{16m^2} \right)^2 + \delta_{\text{I},1}^{\,2}}\,, \label{detunechannel1freq}
\end{align}
with the detuning parameter
\begin{align}
\delta_{\text{I},1} = \frac{1}{2}\left(\frac{e^4 a_1^4}{64m^2} - \frac{e^2 a_2^2 \omega^2}{4m^2} + (E_{2} - E_{-2})(E_0 - 2 \omega) \right)
\label{detunechannel1delta}
\end{align}

\begin{figure}[ht]
\includegraphics[width=0.45\textwidth]{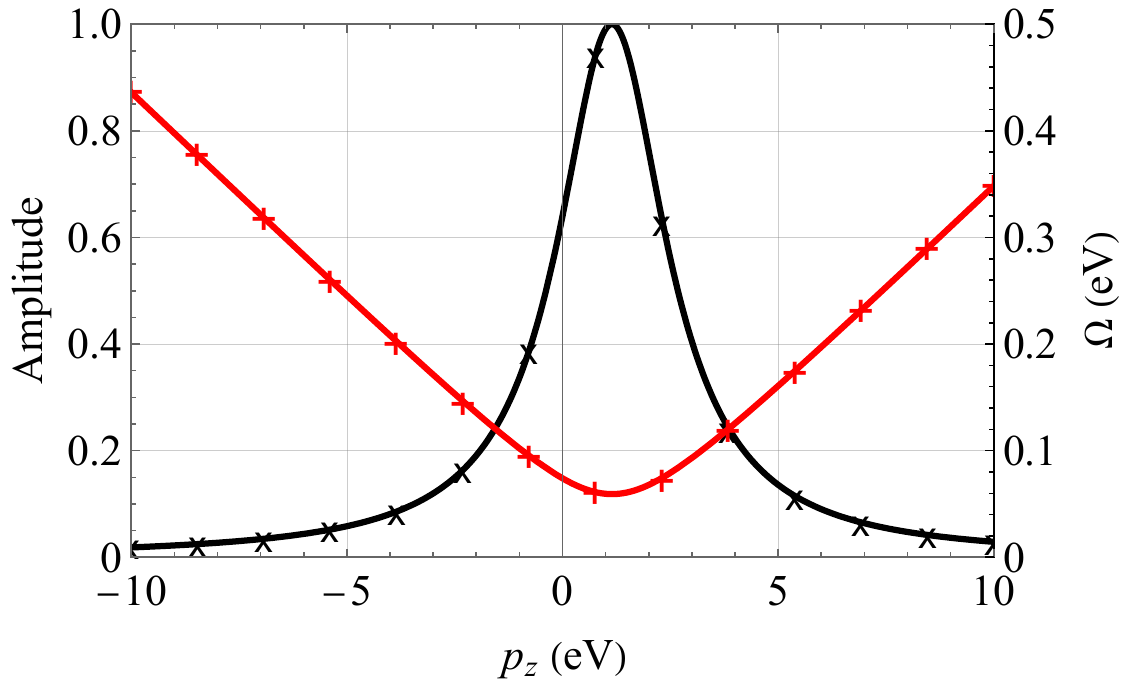}
\caption{The analytical expression \eqref{transitionprobability} for the maximum amplitude of the transition probability $|c_2^\downarrow(t)|^2$ is shown for  channel 1 of subsystem I and the field parameters $|e|a_1 = 10$\,keV, $|e|a_2 = 2.5$\,keV, $\omega = 5$\,keV (black line). The peak of this Lorentz curve is located at $p_z \approx 0.0012$\,keV. Additionally, the corresponding Rabi frequency $\Omega_{\text{I},1}^{(\delta)}$ is shown (red line). The marks (pluses and crosses) indicate numerical simulation results for comparison.}
\label{fig5_LorentzSub1}
\end{figure} 

The resonance behavior of the Rabi oscillation in Eq.~(\ref{transitionprobability}) is shown in Fig.~\ref{fig5_LorentzSub1} for  channel 1 of subsystem~I and the parameters $|e|a_1 = 10$\,keV, $|e|a_2 = 2.5$\,keV, $\omega = 5$\,keV, depicting the maximum amplitude and the oscillation frequency of the transition probability $|c_2^\downarrow(t)|^2$, respectively. They are in very good agreement with the results from numerical simulations based Eq.~(\ref{matrixsub1channel1}) indicated by the markings.

\subsection{Effective Hamiltonian}

In Ref.~\cite{Dellweg-2016} a heuristic model for the detuning behavior of the 3KDE was given by postulating an effective Hamilton operator containing an intrinsic, field-induced detuning [see Eq.~(A1) therein]. From this Hamiltonian, an effective Rabi frequency could be read off describing the detuning and the dynamical behaviour of the system for different offsets $p_z$. With our previous analysis, we are now in a position to derive such an effective Hamiltonian and deduce an analytic solution for the scattering dynamics. We will also see that the resulting effective Rabi frequency coincides with our result in Eq.~\eqref{detunechannel1freq}. 

We start from our system of equation \eqref{generalSystem2ndOrder} 
and introduce a phase transformation by virtue of
\begin{align}
&c_{-2}(t) = e^{-i \frac{1}{2} B_1 t} q_{-2}(t)\,,\ \
c_{2}(t) = e^{-i \frac{1}{2} B_2 t} q_{2}(t)
\end{align}
to remove the damping term. This yields
\begin{align}
&\ddot{q}_{-2}(t) + \Big(A_1 + 	\frac{1}{4} B_1^2 \Big) q_{-2}(t) - e^{-i \frac{1}{2} (B_2 - B_1) t }  \omega_c^2 q_{2}(t) = 0\,,\nonumber\\
&\ddot{q}_{2}(t) + \Big(A_2 + 	\frac{1}{4} B_2^2 \Big) q_{2}(t) - e^{-i \frac{1}{2} (B_1 - B_2) t }  \omega_c^2 q_{-2}(t) = 0\,.
\end{align}
Applying our usual assumption $E_{-2} \approx E_{2}$, so that $B_1 \approx B_2$, the phase factors vanish and the system adopts the form
\begin{align}
\ddot{\vec{q}}(t) = - H^2 \vec{q}(t)\,,
\end{align}
with $\vec{q}(t)= (q_{-2}(t), q_2(t))$.
It can be brought into a hamiltonian form $ i \dot{\vec{q}}(t) = H \vec{q}(t)$ by taking the square root through application of the matrix $Q$ consisting of the eigenvectors of $H^2$ and the associated diagonal matrix $D$ of eigenvalues: 
\begin{align} \label{H}
H = Q \sqrt{D} Q^T
\end{align}
To find the eigenvectors we take advantage of the fact that $H^2$ is symmetric, which allows us to construct $Q$ as a rotation matrix of the form
\begin{align}
Q = \begin{pmatrix}
\cos \theta & -\sin \theta\\
\sin \theta & \cos \theta
\end{pmatrix}
\end{align}
with
\begin{align}
\theta = \frac{1}{2} \arctan \left(\frac{-2\omega_c^2}{A_1 - A_2 + \frac{1}{4}(B_1^2 - B_2^2)}  \right).
\end{align}
For the eigenvalues of $H^2$ we obtain
\begin{align}
d_{\pm} = \frac{1}{2} (A + B)  \pm \sqrt{(A - B)^2 + 4 \omega_c^4}
\end{align}
with $A = A_1 + \frac{1}{4}B_1^2$ and $B = A_2 + \frac{1}{4}B_2^2$
which, after taking the square root, can be Taylor expanded as 
\begin{align}
\sqrt{d_{\pm}} \approx \sqrt{\frac{A+B}{2}} \pm \frac{1}{2} \sqrt{\frac{(A-B)^2 + 4 \omega_c^4}{2(A+B)}}\,.
\end{align}
Insertion into Eq.~\eqref{H} yields
\begin{align}
H &= \frac{A_1 + A_2 + B_1^2}{2 \sqrt{2 (A_1 + A_2)+B_1^2}} \: \openone_2 \notag\\
\notag\\
& + \frac{1}{\sqrt{2(A_1 + A_2) + B_1^2}} \begin{pmatrix}
A_1 & -\omega_c^2 \\
- \omega_c^2 & A_2 
\end{pmatrix}\,.
\end{align}
Since one may neglect any constant term in the Hamiltonian, we can substract $(A_1 + A_2)/2$ on the diagonal to arrive at
\begin{align}
\mathcal{H}_\text{g} = \frac{1}{\sqrt{2(A_1 + A_2) + B_1^2}} \begin{pmatrix}
\frac{1}{2} (A_1-A_2) & - \omega_c^2 \\
- \omega_c^2 & -\frac{1}{2} (A_1-A_2)
\end{pmatrix}  \label{effHamil}
\end{align}
We note here that the denominator of the factor in front of the matrix equals the energy parameter $\varepsilon$ [compare Eq.~(\ref{energyparameter})], which we could approximate in orders of $1/m$. Then the effective Hamiltonian for the channel 1 of subsystem I finally adopts the form
\begin{align}
\mathcal{H}_\text{I,1}= \frac{m}{2\omega(\omega + m)} \begin{pmatrix}
\delta_\text{I,1} & \frac{e^3 a_1^2 a_2 \omega}{16m^2} \\
\frac{e^3 a_1^2 a_2 \omega}{16m^2} & - \delta_\text{I,1}
\end{pmatrix}
\end{align}
with  $\delta_\text{I,1} = \frac{1}{2}\Big[\frac{e^4 a_1^4}{64m^2}  - \frac{e^2 a_2^2 \omega^2}{4m^2} + (E_{2} - E_{-2}) (E_0 - 2\omega) \Big]$. We notice that the latter expression coincides with the detuning parameter in Eq.~(\ref{detunechannel1delta}), as it should be.

\section{Going from individual channels to entire system} \label{GoingToEntireSystem}

In Sec.\,\ref{chapterGeneralSolution} A we have derived the optimum-$p$ formula for a general 3KDE channel and observed that each single laser potential, which couples to an electron state, leads to a shift of the corresponding energy and, this way, to a field-induced detuning in the Rabi oscillation dynamics. Based on this insight we can expect that the addition of further interaction potentials, which arises when merging different channels to a subsystem or different subsystems to the entire system, will combine all their contributions to the energy shift in an additive manner. Collecting all four interaction potentials for the entire 3KDE the corresponding matrix system looks like
\begin{align} \label{entire3KDEsystem}
i \dot{\vec{c}}(t) = \begin{pmatrix}
E_{-2}& \tilde{V}_1 &\tilde{V}_2 &\tilde{W}_1 & \tilde{W}_2 &0\\
\tilde{V}_1^*& E_{0} &0 &0 & 0 &\tilde{W}_2\\
\tilde{V}_2^*& 0 & E_1 &0 &0 &\tilde{W}_1\\
\tilde{W}_1^*& 0 &0 & E_{-1} &0 &\tilde{V}_2\\
\tilde{W}_2^*& 0 &0 &0 & E_0 &\tilde{V}_1\\
0& \tilde{W}_2^* &\tilde{W}_1^* &\tilde{V}_2^* & \tilde{V}_1^* & E_2
\end{pmatrix} \vec{c}(t)\,.
\end{align}
with $\vec{c}(t)= (c_{-2}^\uparrow(t),c_{0}^\uparrow(t),c_{1}(t)^\uparrow,c_{-1}^\downarrow(t),c_{0}^\downarrow(t),c_{2}^\downarrow(t))$. Similarly, the channel Rabi frequencies are expected to be additive, when going from channels over subsystems to the entire system. This is indeed the case, as we will verify in the following by simulations.

\subsection{Optimum-p formula}

\begin{figure}[ht]
\includegraphics[width=0.45\textwidth]{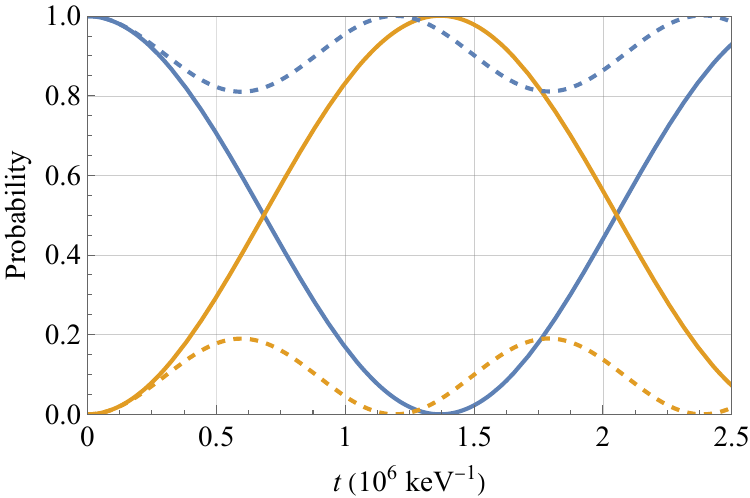}
\caption{Rabi oscillation dynamics of subsystem I in Eq.~(\ref{matrixSub1}). Dashed lines indicate the case where $p_z =0$ and solid lines the case where $p_z\approx p_{\text{I},1} + p_{\text{I},2} \approx 0.121$\,eV is given by the corresponding optimum-$p$ formula. The field parameters are chosen as $|e|a_1 = 10$\,keV, $|e|a_2 = 4.9$\,keV, and $\omega = 5$\,keV.} \label{fig6_Sub1}
\end{figure}

According to Eqs.~\eqref{pformelSub1Channel1} and \eqref{p_g} the optimum-$p$ formulas for channels 1 and 2 of subsystems I and II read
\begin{align}
&p_{\text{I},1} \approx \frac{e^4 a_1^4 - 16 e^2 a_2^2 \omega^2}{512 m \omega^2} \approx p_{\text{I},2}\,,\nonumber\\
&p_{\text{II},1} \approx \frac{-e^4 a_1^2 a_2^2 -  e^2 a_1^2 \omega^2}{64 m \omega^2} \approx p_{\text{II},2}\,.
\end{align}
We recall that the interactios potentials (e.g., $\tilde{V}_1$ and $\tilde{W}_2$ for subsystem I) act in channel 2 in inverse order as compared with channel 1, so that the roles of $U_1$ and $U_2$ in Eq.~\eqref{p_g} are exchanged. Since in this equation the frequency $\omega_1$ is simultaneously to be replaced by $\omega_2=-\omega_1$, the optimum momentum for channel 2 comes out (approximately) equal to channel 1. While $p_{\text{I},1}$ is able to compensate the detuning in channel 1 of subsystem I [see Eq.~\eqref{matrixsub1channel1}] and accordingly for $p_{\text{I},2}$, the whole subsystem I contains twice as many potentials as each channel alone [see Eq.~\eqref{matrixSub1}], so that the combined momentum shift $p_\text{I} \approx p_{\text{I},1} + p_{\text{I},2}$ is required to compensate the field dressing in the subsystem. This conclusion is confirmed by the numerical simulation shown in Fig.~\ref{fig6_Sub1}. It holds correspondingly for subsystem II, as well, whose optimum momentum is given by $p_\text{II} \approx p_{\text{II},1} + p_{\text{II},2}$. 

\begin{figure}[ht]
\includegraphics[width=0.45\textwidth]{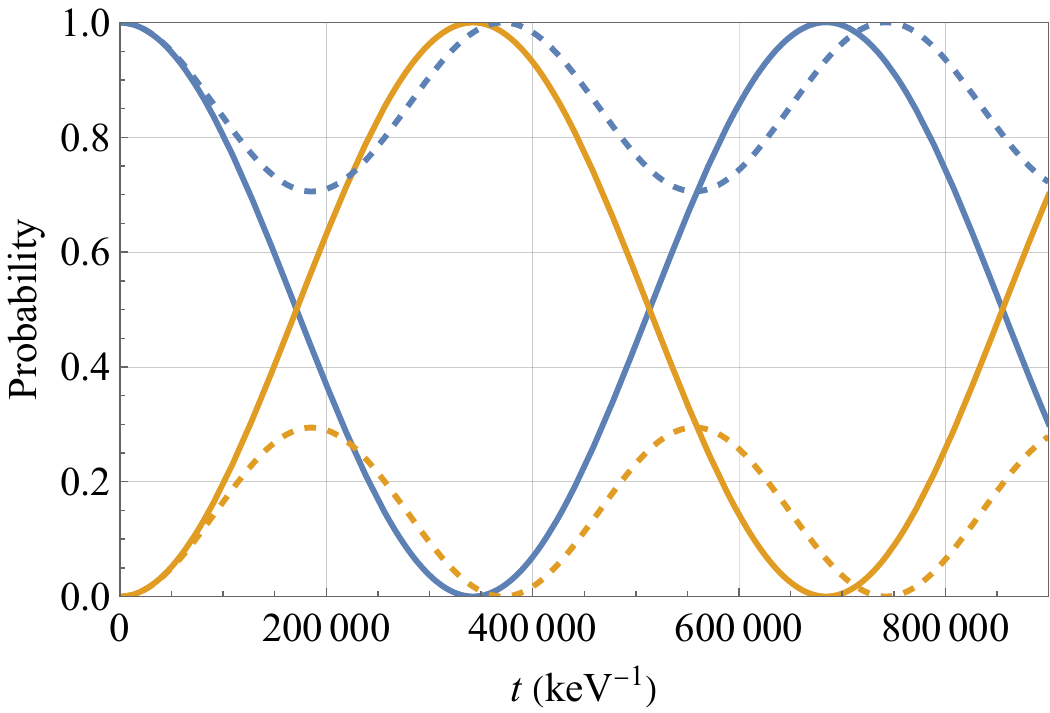}
\caption{Rabi oscillation dynamics of the entire 3KDE model system in Eq.~(\ref{entire3KDEsystem}). Dashed lines indicate the case $p_z=0$ and solid lines the case where $p_z \approx 0.363$\,eV is given by the optimum-$p$ formula in Eq.~(\ref{3KDEpformula}). The field parameters are $|e|a_1 = 10$\,keV, $|e|a_2 = 4.9$\,keV, and $\omega = 5$\,keV.}
\label{fig7_3KDE}
\end{figure}

Along the same line of argument, we obtain the optimum-$p$ formula for the entire system (\ref{entire3KDEsystem}) by virtue of
\begin{align}
p_\text{3KDE} &\approx  p_\text{I} + p_\text{II} \nonumber\\
&\approx  \frac{e^4 a_1^4 - 16 e^2 a_2^2 \omega^2}{256 m \omega^2} +  \frac{-e^4 a_1^2 a_2^2 +  e^2 a_1^2 \omega^2}{32 m \omega^2} \label{3KDEpformula}
\end{align}
The effectiveness of this expression to cure the field-induced detuning is demonstrated in Fig.~\ref{fig7_3KDE} for the parameters $|e|a_1 = 10$\,keV, $|e|a_2 = 4.9$\,keV, and $\omega = 5$\,keV. A momentum offset of $p_z \approx 0.363$\,eV leads to a fully developed Rabi oscillation, whereas for $p_z = 0$ the maximum scattering probability drops down to about $30\%$, even though the applied field configuration differs only slightly from the case of optimal laser parameters (which would require $|e|a_2 = 5$\,keV).

Referring to Eq.~\eqref{allgemeineFrequenz}, the Rabi frequencies for the different channels read
\begin{align}
&\Omega_{\text{I}, 1/2} 
\approx \frac{m}{\omega ( \omega \pm m)} 
\sqrt{ \omega_c^4 + {\delta^{\,2}_{\text{I},1/2}}}\,, \nonumber\\
&\Omega_{\text{II}, 1/2}
\approx \frac{4m}{\omega ( 3\omega \mp 2m)} 
\sqrt{ \omega_c^4 + {\delta^{\,2}_{\text{II},1/2}}}\,,
\end{align}
with the detuning parameters
\begin{align}
&\delta_{\text{I}, 1/2} = \frac{1}{2}\left(\pm \frac{a_1^4}{64m^2} \mp \frac{a_2^2 \omega^2}{4m^2} + (E_{2} - E_{-2})(E_0 \mp 2 \omega) \right),\nonumber\\
&\delta_{\text{II}, 1/2} = \frac{1}{2}\left(\pm  \frac{a_1^2 a_2^2}{16m^2} \mp \frac{a_1^2 \omega^2}{16m^2} + (E_{2} - E_{-2})(E_1 \pm  \omega) \right).
\end{align}

Analogously to the optimum electron momentum, the Rabi frequency for subsystem I is obtained as the sum of the channel Rabi frequencies, $\Omega_\text{I} \approx \Omega_{\text{I},1} + \Omega_{\text{I},2}$, and correspondingly for $\Omega_\text{II}$ of subsystem II. 

In the case, when there is no detuning, one has
\begin{align} \label{Omega_I_II}
\Omega_\text{I} \approx -\frac{2m\omega_c^2}{\omega^2 - m^2}\ \ \ \text{and}\ \ \
\Omega_\text{II} \approx -\frac{6m\omega_c^2}{\frac{9}{4}\omega^2 - m^2}\,.
\end{align}
We note that the subsystem~II oscillates about thrice as fast as subsystem~I, since $\Omega_\text{II}\approx 3\Omega_\text{I}$. Furthermore it is worth mentioning that the terms $\Omega_\text{I}$ and $\Omega_\text{II}$ of Eq.~\eqref{Omega_I_II} also arise in the perturbative treatment of the 3KDE, where they are added to derive the Rabi frequency of the full 3KDE process for the case without detuning (see Eq.~(15) in Ref.~\cite{Dellweg-2016}). In agreement with this treatment, we obtain here
\begin{align} \label{Omega_3KDE}
\Omega_\text{3KDE} \approx \Omega_\text{I} + \Omega_\text{II} \approx -\frac{e^3 a_1^2 a_2 \omega}{2m^3}\,.
\end{align}

The oscillation frequency found in the numerical simulations of Fig.~\ref{fig7_3KDE} coincides with this expression. We point out that the resonant Rabi frequency of the full KDE process is substantially smaller than the Rabi frequency of a single channel. For example, from Eq.~\eqref{sub1channel1freq} follows $\Omega_\text{3KDE}/\Omega_\text{I,1}^{(0)}\approx 8\omega/m \ll 1$, which also becomes apparent by comparing Figs.~\ref{fig3_Sub1Channel1} and \ref{fig7_3KDE}. In comparison with the Rabi frequency of subsystem I in Fig.~\ref{fig6_Sub1}, the full resonant Rabi frequency $\Omega_\text{3KDE}$ is four times larger.

The total transition probability for the entire 3KDE model system can be written as
\begin{align}
|c^\downarrow_{2,\text{3KDE}}(t)|^2  
= \left( \frac{\Omega_\text{3KDE}}{ \Omega_\text{3KDE}^{(\delta)} } \right)^{\! 2} \sin^2 \left(\tfrac{1}{2} \Omega_\text{3KDE}^{(\delta)}  t \right) \label{matthias3KDEgesamt}
\end{align}
with the Rabi frequency including the detuning
\begin{align}
\Omega_\text{3KDE}^{(\delta)} =  \sqrt{  \Omega_\text{3KDE}^{2}+ \left(\Delta_\text{3KDE} - \frac{4\omega p_z}{m} \right)^{\!2} }
\end{align}
where  $\Delta_\text{3KDE} = (4\omega/m)p_\text{3KDE}$. 

\begin{figure}[ht]
\includegraphics[width=0.45\textwidth]{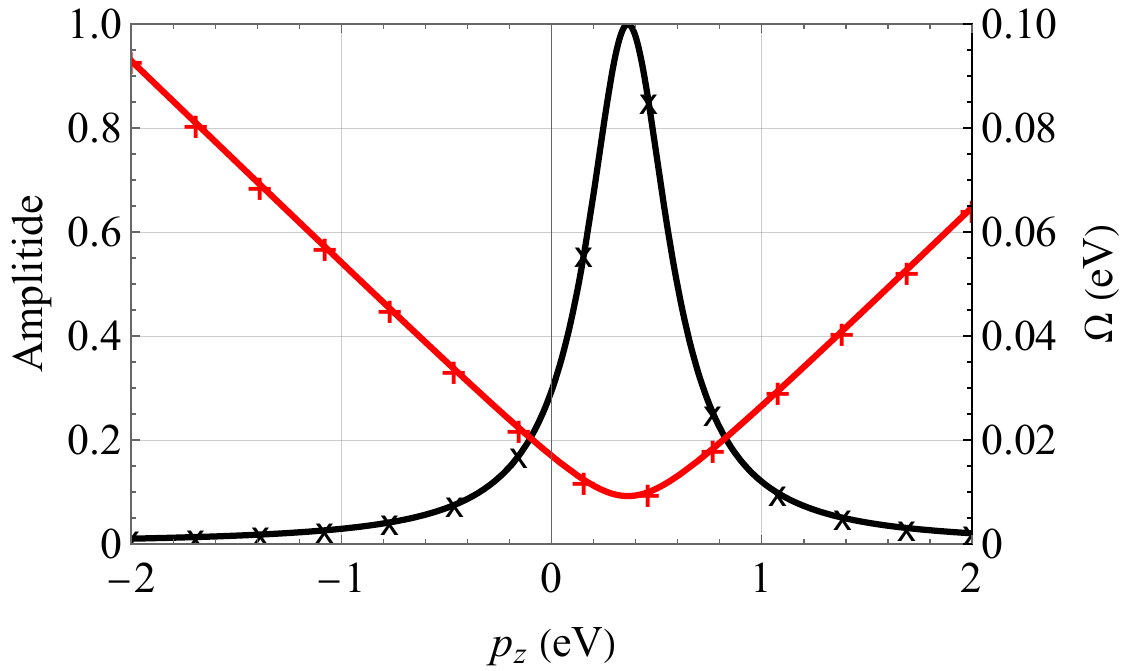}
\caption{Lorentz curve of the transition probability $|c_2^\downarrow(t)|^2$ for the entire 3KDE system (black line). Additionally, the corresponding Rabi frequency $\Omega_\text{3KDE}^{(\delta)}$ is shown (red line), and numerical simulation results are indicated by markings. The field parameters are the same as in Fig.~\ref{fig7_3KDE}.}
\label{fig8_Lorentz3KDE}
\end{figure}

The maximum amplitude and Rabi frequency of the scattering probability in Eq.~\eqref{matthias3KDEgesamt} are illustrated in Fig.~\ref{fig8_Lorentz3KDE} for the field parameters $|e|a_1 = 10$\,keV, $|e|a_2 = 4.9$\,keV, and $\omega = 5$\,keV. The Lorentz curve of the maximum amplitude, which exhibits a peak at $p_z \approx 0.363$\,eV, is much more narrow than the one in Fig.~\ref{fig5_LorentzSub1} for the channel system. This is because the Rabi frequency of the entire system is smaller by a relative factor $8\omega/m$ than in the case of a single channel. Thus, the entire system reacts much more sensitively to deviations from the optimal case.

\section{Treatment of 4KDE}\label{Treatment4KDE}

After having analyzed the spin-dependent 3KDE in detail, in the following we briefly consider the 4KDE. The mathematical descriptions of both KDE versions possess close similarities, since they involve two transition potentials to leading order. For the 4KDE, they read 
\begin{align}
&V_{1} = \frac{e^2 a_1^2}{8m} e^{2 i \omega t} \sum_n \ket{n}\bra{n+2}\,,\nonumber\\
&V^{\text{4KDE}}_{2} = \frac{e^2 a_1 a_2}{4m} e^{-2 i \omega t} \sum_n \ket{n}\bra{n+4}\,.
\end{align}
From this we can construct a minimal model system in the form of the matrix equation 
\begin{align} \label{matrix4KDE}
i \dot{\vec{c}}(t) = \begin{pmatrix}
E_{-3}& \tilde{V}_{1} &\tilde{V}_2 &0\\
\tilde{V}_1^*& E_{-1} &0 &\tilde{V}_2\\
\tilde{V}_2^*& 0 &E_{1} &\tilde{V}_1\\
0& \tilde{V}_2^* &\tilde{V}_1^* &E_3
\end{pmatrix} \vec{c}(t)
\end{align}
with $\vec{c}(t) = (c_{-3}(t),c_{-1}(t),c_{1}(t),c_3(t))$ and
\begin{align}
\tilde{V}_1 = \frac{e^2a_1^2}{8m} e^{2i \omega t}, \quad 
\tilde{V}_2 = \frac{e^2a_1 a_2 }{4m} e^{-  2 \omega t}\,.
\end{align}
We emphasize that the interaction potentials do not couple to the electron spin. Thus, to leading order, the 4KDE is a spin-preserving process. Therefore, we have omitted the spin-superscripts in the coefficients $c_n(t)$. If the electron is incident with spin up, this spin projection will remain unchanged throughout the transition. It is a general feature of $\mathcal{N}$-KDE processes that they are spin-preserving to leading order when $\mathcal{N}$ is even. This is because the interaction is governed by $\vec{A}^2$ terms.

Despite this physically relevant difference, the matrix in Eq.~\eqref{matrix4KDE} closely resembles the corresponding matrix of subsystem I of the 3KDE [see Eq.~\eqref{matrixSub1}]. Consequently, one can treat the 4KDE like the 3KDE but without the need of handling two separate subsystems. 

\subsection{Rabi Frequency of the 4KDE}

For the 4KDE without detuning, we can derive the Rabi frequency analogously to Ref.~\cite{Dellweg-2016} using time-dependent perturbation theory. With the free propagator
\begin{align}
U_0(t-t') = \sum_n e^{-i E_n(t-t')} \ket{n}\bra{n}
\end{align}
we can consider the Dyson series
\begin{align}
\braket{3|U(T)|-3} &\approx - \int_0^T dt' \int_0^{t'} dt'' \bra{3}U_0(T-t') \notag \\
&\quad \times [V_1(t') + V_2(t')] U_0(t'-t'') \notag\\
&\quad \times [V_1(t'') + V_2(t'')] U_0(t'') \ket{-3} \notag \\
&= - \frac{e^4 a_1^3 a_2}{32m^2} (I_+ + I_-).
\end{align}
The integrals are
\begin{align}
I_\pm &= e^{-i E_3 T} \int_0^T dt' e^{-i(-E_3 + E_{\pm 1} \mp 2\omega)t'} \notag\\
&\quad \times  \int_0^{t'} dt''  e^{-i(E_{-3} - E_{\pm 1} \pm 2 \omega)t''} \notag\\
&\approx i \frac{T e^{-i E_3 T}}{E_{-3} - E_{\pm 1} \pm 2 \omega}\,, 
\end{align}
where $T$ denotes the interaction time. The transition amplitude then becomes
\begin{align}
\braket{3|U(T)|-3} &\approx i \frac{e^4 a_1^3 a_2}{16 m^3} T  e^{-i E_3 T} \frac{1}{1-\frac{4\omega^2}{m^2}} \nonumber\\
&\approx i \frac{e^4 a_1^3 a_2}{16m^3}T e^{-i E_3 T}\,,
\end{align}
from which we can read off the resonant Rabi frequency for the 4KDE as
\begin{align}
\Omega_\text{4KDE} \approx \frac{e^4 a_1^3 a_2}{8m^3}. \label{4KDEfrequency}
\end{align}
We remark that the 4KDE Rabi frequency is of the order $\mathcal{O}\big(m^{-3}\big)$ just like the 3KDE Rabi frequency of Eq.~\eqref{Omega_3KDE}. When higher orders in the interaction with the laser fields are taken into account, also spin-flipping transitions may occur in the 4KDE which are of the order $\mathcal{O}\big(m^{-4}\big)$. They have been studied in Ref.~\cite{4KDE}, where the dominant spin-preserving transition has been disregarded, though.

\subsection{Optimum-p formula for the 4KDE}

\begin{figure}[ht]
\includegraphics[width=0.45\textwidth]{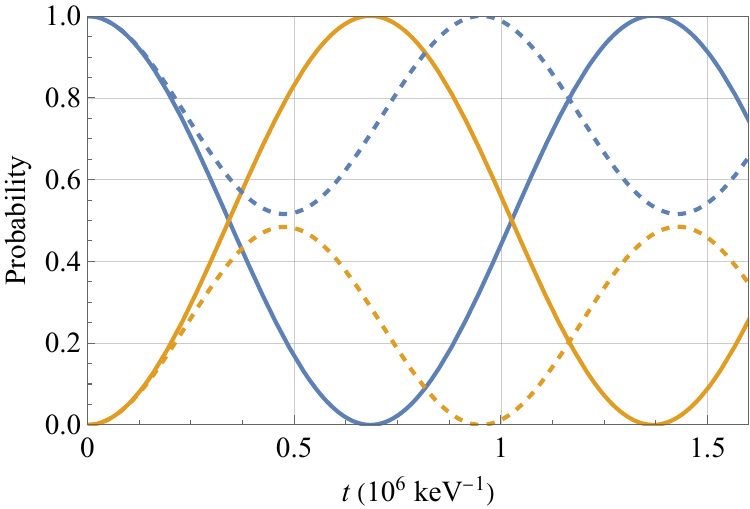}
\caption{Compensated detuning for the 4KDE (solid lines) by utilizing the optimum-$p$ formula (\ref{4KDEpformula}) for the parameters $|e|a_1 = 10$\,keV, $|e|a_2 = 4.9$\,keV, $\omega = 5$\,keV with $p_z \approx 0.081$\,eV. The detuned case for $p_z = 0$ is indicated by the dashed lines.}
\label{fig9_4KDE}
\end{figure}

From the perturbation theory in Sec.\,\ref{Treatment4KDE} A and from the transition potentials in Eq.~\eqref{matrix4KDE} we see that the 4KDE system consists of two separate channels being
\begin{align}
&\text{channel 1:} \quad c_{-3} \overset{V_{1}}{\longrightarrow} c_{-1} \overset{V_{2}}{\longrightarrow} c_{3}\,,\nonumber\\
&\text{channel 2:} \quad c_{-3} \overset{V_{2}}{\longrightarrow} c_{1} \overset{V_{1}}{\longrightarrow} c_{3}\,.
\end{align}
As before, we can analyze each channel separately, with the matrix of channel 1 reading
\begin{align}
i \dot{\vec{c}}(t) = \begin{pmatrix}
E_{-3}& \tilde{V}_{1}  &0\\
\tilde{V}_1^*& E_{-1}  &\tilde{V}_2\\
0& \tilde{V}_2^*  &E_3
\end{pmatrix} \vec{c}(t)\,.
\end{align}
By inserting the first and last equation into the middle one, we can obtain a system of two coupled differential equations of second order:
\begin{align}
&\ddot{c}_{-3}(t) + i (E_{-3} + E_{-1} - 2 \omega)  \dot{c}_{-3}(t) \notag \\
&+ \left(\frac{e^4a_1^4}{64m^2} - E_{-3}(E_{-1} - 2 \omega)\right) c_{-3}(t)  + \frac{e^4 a_1^3 a_2 }{32m^2} c_{3}(t) = 0\,,\nonumber\\
&\ddot{c}_{3}(t) + i (E_{3} + E_{-1}  - 2 \omega)  \dot{c}_{3}(t) \notag \\
&+ \left(\frac{e^4 a_1^2 a_2^2}{16m^2} - E_{3}(E_{-1} - 2 \omega)\right) c_{3}(t)  + \frac{e^4 a_1^3 a_2}{32m^2} c_{-3}(t) = 0
\end{align}
Both equations are coupled by a hook term similiar to the 3KDE case involving the product of both transition potentials. As expected, the structure of these equations is identical to those for the 3KDE in Eq.~\eqref{sub1channel1secondOrderEq}. Since $E_{-3}\approx E_3$, they become symmetric if
\begin{align*}
\frac{e^4 a_1^4}{64m^2} - E_{-3}(E_{-1} - 2 \omega)\overset{!}{=} \frac{e^4 a_1^2 a_2^2}{16m^2} - E_{3}(E_{-1} - 2 \omega)
\end{align*}
which results in the optimum-$p$ formula for correcting the detuning in channel 1 as
\begin{align}
p^{\text{4KDE}}_1 = - \frac{m}{6 \omega} \frac{e^4 (a_1^4 - 4 a_1^2 a_2^2)}{64m^2(E_{-1} - 2 \omega)} \approx \frac{e^4 (a_1^4 - 4a_1^2 a_2^2)}{768  m \omega^2}\,.
\end{align}
We can read off the relation for optimal laser parameters, for which no field-induced detuning arises, as
\begin{align}
a_1^2 - 4 a_2^2 = 0.
\end{align}
When considering channel 2, the transition potentials switch their positions and the exchanged photon energy has opposite sign, leading to the optimum momentum 
\begin{align}
p^{\text{4KDE}}_{2} = - \frac{m}{6 \omega} \frac{e^4 (a_1^4 - 4 a_1^2 a_2^2)}{64m^2(E_{1} + 2 \omega)} \approx p^{\text{4KDE}}_{1}\,.
\end{align}
As a result, the optimum-$p$ formula for the entire 4KDE model system of Eq.~\eqref{matrix4KDE} is given by 
\begin{align}
p^{\text{4KDE}} \approx p^{\text{4KDE}}_{1} + p^{\text{4KDE}}_{2} \approx
\frac{e^4 (a_1^4 - 4a_1^2 a_2^2)}{384 m \omega^2}\,. \label{4KDEpformula}
\end{align}
The compensating effect of the optimum-$p$ formula is shown in Fig.~\ref{fig9_4KDE} where $p_z \approx 0.081$\,eV leads to a fully developed Rabi oscillation.

\section{Conclusion and Outlook}\label{ConclusionOutlook}

Nonrelativistic Kapitza-Dirac scattering of electrons from counterpropagating bichromatic laser waves was studied in the resonant Bragg regime, taking the electron spin into account. We developed minimal model systems preserving the characteristic properties of the process, which allowed us to analyze quantitatively the intrinsic field-induced detuning in the Rabi oscillation dynamics between the incident and scattered electron states. 

To this end, we decomposed the scattering process into separate channels and showed that each of them behaves like a system of two coupled oscillators, whose eigenmodes are shifted due to the coupling via the laser fields, so that the field-induced detuning can be seen as an avoided-crossing phenomenon. From this we derived suitable adjustments of the incident electron momentum to compensate the detuning for each channel, analyzed the resulting detuned Rabi frequencies, and found effective Hamiltonians governing the channel dynamics. These findings enabled us to describe the Rabi oscillation dynamics between incident and scattered electron state fully analytically and to merge our channel results to partially restore the original dimensionality of the problem. 

As a major final step, which will be subject to forthcoming studies, it remains to construct a proper description of the full, infinitely dimensional system of equations from our minimal-model results for the KDE. This step is challenging because it requires to account for a number of further dynamical elements such as ``backward loops'' of the form $c_{-2}^\uparrow \overset{V_1^*}{\longrightarrow} c_{-4}^\uparrow \overset{V_1}{\longrightarrow} c_{-2}^\uparrow$, which can lead to additional energy shifts that need to be compensated by an adjusted momentum offset. The physical insights of the present study have deepened our understanding of the resonance features of the KDE and pave the way towards a full solution of its detuning problem.


\end{document}